\documentstyle[prb,aps,psfig,epsf,epsfig,twocolumn]{revtex}
\newcommand {\be}{\begin{equation}}
\newcommand {\ee}{\end{equation}}

\tighten

\begin{document}

\title{\bf Model of Hydrophobic Attraction in Two and Three Dimensions}
\author{G.T. Barkema and B. Widom~\cite{widomadd}}
\address{Instituut voor Theoretische Fysica, Universiteit Utrecht,\\
Princetonplein 5, 3584 CC Utrecht, The Netherlands}
\date{March 22, 2000}

\maketitle

\begin{abstract}
An earlier one-dimensional lattice model of hydrophobic attraction is
extended to two and three dimensions and studied by Monte Carlo
simulation. The solvent-mediated contribution to the potential of mean
force between hydrophobic solute molecules and the solubility of the
solute are determined.  As in the earlier model, an inverse relation is
observed between the strength and range of the hydrophobic attraction.
The mean force no longer varies monotonically with distance, as it does
in one dimension, but has some oscillations, reflecting the greater
geometrical complexity of the lattice in the higher dimensions. In
addition to the strong attraction at short distances, there is now also
a local minimum in the potential of depth about $kT$ at a distance of
three lattice spacings in two dimensions and one of depth about $2kT$
at a distance of two lattice spacings in three dimensions. The
solubility of the solute is found to decrease with increasing
temperature at low temperatures, which is another signature of the
hydrophobic effect and also agrees with what had been found in the
one-dimensional model.
\end{abstract}

\section{Introduction}

A one-dimensional lattice model of hydrophobic attraction has recently
been studied~\cite{kolomeisky99}. The model incorporates what is
believed to be the basic mechanism of the hydrophobic effect; viz.,
that the accommodation of the solute in the solvent is energetically
favorable but sufficiently unfavorable entropically as to result in an
increase in free energy. This increase is smaller when the hydrophobes
are close together than when they are widely separated, resulting in a
net solvent-mediated attraction between them.

The unfavorable entropy of accommodating the solute is achieved in the
model by allowing each solvent molecule to have a large number $q$ of
possible orientations but requiring two neighboring solvent molecules
to be both in a special one of those $q$ states if they are to
accommodate a hydrophobic solute in the interstitial site between them.
That circumstance in which two neighboring solvent molecules are both in
that special orientation is taken to be energetically favorable by an
amount previously called~\cite{kolomeisky99} $u-w$. The condition of
hydrophobicity at temperature $T$ was that $kT\ln(q-1)>u-w$, with $k$
Boltzmann's constant. To model the hydrophobic effect realistically,
$(u-w)/k$ was taken to be $3000$K and $q$ was taken to be 110,000.

It was then a property of the model that forcing the solvent to
accommodate a hydrophobic solute greatly restricted (to one state out
of $q$) the possible orientations of the solvent molecules neighboring
the accommodated solute. Such restriction of solvent orientation by the
solute is consistent with experimental measurements of NMR relaxation
times, which show that the reorientation times of water molecules that
neighbor a solute with alkyl groups are about twice as long as in pure
water, and that there is an increased activation energy for the
reorientation in the presence of the solute over that in pure
water.\cite{hertz64}

It was a striking feature of the one-dimensional model that the range
and strength of the solvent-mediated attraction between solutes were
inversely related to each other; at low temperatures the attraction was
weak but long-ranged and at high temperatures it was strong but
short-ranged. One of the motivations of the present work was to see if
that inverse relation, which is of potential importance in the
interpretation of experiment, persists in higher dimensions, and it is
indeed found to be so.

It was also found earlier in the one-dimensional model that the
solubility of the hydrophobe decreases with increasing temperature,
which is consistent with the increasing hydrophobicity that manifests
itself in the increasing strength of the hydrophobic attraction. This
temperature dependence of the solubility, too, is found here to persist
in the two- and three-dimensional models. It is also found, as in the
one-dimensional model, that if the energy of interaction of the
hydrophobic solute with its solvent neighbors (earlier called $v$) is
greater than the magnitude $u-w>0$ of the favorable energy of solvent
alignment, then the solubility, after first decreasing with increasing
temperature, reaches a minimum and then increases.

The two- and three-dimensional versions of the model are defined in the
following section. Also in Section \ref{sec:models} are the basic
formulas that relate the solvent-mediated part of the potential of mean
force and the solubility to the quantities that are measured in the
simulations. The details of the simulations are in Section
\ref{sec:simulation}, where the results are also displayed. These are
discussed and summarized in the concluding Section \ref{sec:summary}.

\section{The two- and three-dimensional models}
\label{sec:models}

The two-dimensional model is pictured in Fig. 1. At each site of a
square lattice there is a solvent molecule that may be in any of a
large number, $q$, of different states (orientations). Solute molecules
may only be accommodated at the interstitial sites between neighboring
pairs of solvent molecules, but only if that site is not already
occupied and then only if the two solvent molecules in question are
both in a special one of the $q$ states, called state 1 in what
follows. The figure shows three solute molecules so accommodated.

\begin{center}
\begin{figure}
\epsfxsize 5cm
\epsfbox{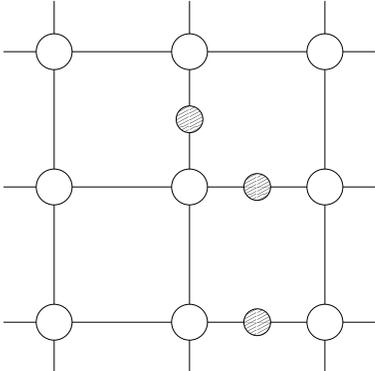}
\caption{Two-dimensional model on a square lattice. There is a solvent
molecule (open circle) at each lattice site; solute molecules (shaded
circles) may only occupy interstitial sites.}
\label{modeltwo}
\end{figure}
\end{center}

The three-dimensional model is defined similarly, and also on a lattice
of coordination number 4, as in two dimensions. This
three-dimensional lattice is pictured in Fig. 2. It is obtained by
systematically removing two bonds from each vertex of a simple cubic
lattice, in such a way that from each site $(i,j,k)$ there remains a
bond to the neighboring sites at $(i+1,j,k)$ and  $(i-1,j,k)$, as well as
to those at  $(i,j-1,k)$ and  $(i,j,k-1)$ if $i+j+k$ is even but to
those at $(i,j+1,k)$ and $(i,j,k+1)$ instead if $i+j+k$ is odd. The
lattice is topologically (not metrically) equivalent to the tetrahedral
diamond lattice, which is also the lattice of oxygen atoms in the metastable
phase of ice called ice I$_c$.

\begin{figure}
\epsfxsize 8cm
\epsfbox{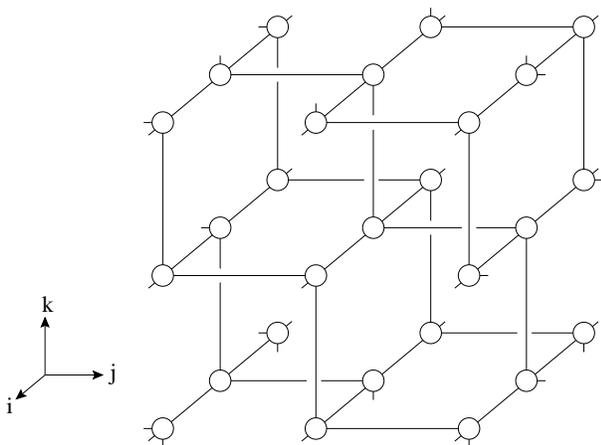}
\caption{Four-coordinate lattice in three dimensions obtained by
systematically removing two bonds from each vertex of a simple cubic
lattice. Circles mark the lattice sites, of which 27 are shown, in three
horizontal $(i,j)$ or vertical $(j,k$ or $k,i)$ planes of nine sites each.}
\label{modelthree}
\end{figure}

In both the two- and three-dimensional versions of the model only
neighboring solvent molecules interact with each other, which they do
with energy $w$ if both are in the special state 1 and with energy
$u>w$ otherwise. The only solvent molecules with which a solute
molecule interacts are its immediate neighbors; the sum of its energies
of interaction with them is $v$.

Solute molecules may interact with each other but that is irrelevant
here; the solute-solute interaction does not enter the expression for
the {\it solvent-mediated part} of the potential of mean force
between pairs of solutes in the infinitely dilute
solution~\cite{kolomeisky99}, nor is it relevant for the solubility of
the solute, which is so slight that solute-solute interactions are
negligible. The solute-solvent interaction parameter $v$ is also absent
from the expression for the potential of mean force in the infinitely
dilute solution but it is a significant parameter in the
solubility.\cite{kolomeisky99}

Because the coordination number of the lattices considered here is 4,
rather than 2 as in one dimension, the values of the parameters $u-w$
and $q$ required to model the hydrophobic effect conveniently but still
realistically over the temperature range 275-375K are both much
smaller than in one dimension. Here, in both the two- and
three-dimensional models, they are taken to be $(u-w)/k=1000$K instead
of 3000K and $q=1100$ instead of 110,000. It is still the case, as in
the one-dimensional model, that $kT\ln(q-1)>u-w$ over the relevant
temperature range, and that the extent to which $kT\ln(q-1)$ exceeds
$u-w$ increases with increasing temperature --- which are the respective
conditions required for the hydrophobic effect to manifest itself and
to do so with increasing strength as the temperature increases.

Let $p_{11}$ be the probability that in the pure solvent both molecules
of a given neighboring pair will be found in the special state 1; and
let $p(r)$ be the probability that the molecules at two such pairs of
sites will all be found to be in the special state 1 when the
corresponding two interstitial sites, one associated with each pair,
are separated by the metrical distance $r$ (measured in units of the
lattice spacing). In the two-dimensional lattice of Fig. 1 the values
of $r$ that occur are $r=\sqrt{2}/2=0.71; 1; \sqrt{2}=1.41;
\sqrt{5/2}=1.58$; etc.  In the three-dimensional lattice of Fig. 2 the
values of $r$ that occur are those and also $r=\sqrt{3/2}=1.22$, etc.

Let $W(r)$ be the solvent-mediated part of the potential of mean force
between two solute molecules a distance $r$ apart, in the limit of
infinite dilution. This is the potential of mean force from which the
direct interaction between the two solute molecules themselves has been
subtracted, and is the part of the mean-force potential that is of
interest here. Then, as in the one-dimensional
model,\cite{kolomeisky99} $W(r)$ is given in terms of $p_{11}$ and
$p(r)$ by
\begin{equation}
W(r)=-kT\ln\left[p(r)/p_{11}^2\right].
\label{eq:W}
\end{equation}
The quantities $p_{11}$ and $p(r)$ are what are determined in the
simulations and $W(r)$ is then obtained from (1). It is a property of
the solvent alone, and so depends only on the parameters $u-w$ and $q$
but not on $v$.

The solubility is defined as in the earlier one-dimensional
model.\cite{kolomeisky99} One imagines a hypothetical ideal gas of
pure solute, of number density $\rho_g$, in osmotic
equilibrium with the saturated solution; {\it i.e.}, having a
thermodynamic activity equal to that of the solute in the solution.
Then if $\rho_s$ is the number density of the solute in the
solution, the solubility $\Sigma$ is here defined as the dimensionless
ratio
\begin{equation}
\Sigma=\rho_s/\rho_g.
\label{eq:Sigma}
\end{equation}
If the solubility is low so that the saturated solution is very dilute,
this $\Sigma$ is the Ostwald absorption coefficient and is given
by\cite{kolomeisky99}
\begin{equation}
\Sigma=p_{11}{\rm e}^{-v/kT}.
\label{eq:solubility}
\end{equation}
Thus, $\Sigma$ is obtained once $p_{11}$ is calculated and a value of
the solute-solvent interaction energy $v$ is specified.

\section{Simulations}
\label{sec:simulation}

Properties of the two- and three-dimensional models presented in the
previous Section are explored by means of Monte Carlo simulations. In
these simulations, one solvent molecule is placed on each lattice
site.  For each molecule, we keep track of whether its state is the
special state 1, or not.  Solute molecules are not present in the Monte
Carlo simulations, since the aim of the simulations is to numerically
estimate the quantities $p_{11}$ and $p(r)$, from which the
solvent-mediated part of the potential of mean force can be computed
using (\ref{eq:W}).

The dynamics of the models are not specified, and we are therefore free
to choose any dynamics that yield the correct equilibrium properties.
In our Monte Carlo simulations, the dynamics consist of two processes:
1) each solvent molecule that is in its special state 1 leaves this
special state with a rate of $q-1$ per time unit;
2) each solvent molecule that is in a different state enters its
special state with a rate of $\alpha^n$ per time unit, where
$\alpha=\exp((u-w)/kT)$ and $n$ is the number of nearest-neighbor sites
that are occupied with a solvent molecule in its special state.
An efficient implementation of these dynamics is obtained with
a BKL-scheme, after Bortz, Kalos and Lebowitz~\cite{bkl}. 

The simulations are performed on a $500\times 500$ lattice for the
two-dimensional model, and a $50\times 50\times 50$ lattice for the
three-dimensional model, both with periodic boundary conditions in
order to reduce finite-size effects. In the initial configuration, no
molecule is in its special state. This configuration is then
thermalized for 0.1 time units.  After thermalization, $p_{11}$ and
$p(r)$ are measured every $10^{-3}$ time unit, up to a total time of
$10^3$ time units, and averaged. In a simulation of this length and for
these system sizes, the two-dimensional and three-dimensional model
undergo roughly $5.5\cdot 10^8$ and $2.7\cdot 10^8$ local changes,
respectively. This procedure is repeated 10 times for each of the
temperatures $T$=275, 300, 325, 350, and 375K, and for the
three-dimensional model also at a temperature of 250K at which the
system is probably metastable, the stable regime being one where almost
all atoms are in their special state.

The values for the quantity $p_{11}$ are presented in Table
\ref{tab:p11}, for the two- and three-dimensional models. This quantity
is decreasing with increasing temperature. The quantity $p_{11}$ is one
of the two factors determining the solubility as defined in equation
(\ref{eq:solubility}), the other being $\exp(-v/kT)$. While this latter
factor is monotonically increasing with increasing temperature for
$v>0$, our data show that $p_{11}$ is monotonically decreasing with
increasing temperature. For some choices of $v$, the combination of
these two factors leads to a solubility that is initially decreasing
with increasing temperature (a signature of the hydrophobic effect),
but later increasing.

The solvent-mediated part of the potential of mean force $W(r)$ is
plotted as a function of the distance $r$ in Fig. 3 for the
two-dimensional model, and in Fig. 4 for the three-dimensional model.
In the latter figure, the curve obtained at a temperature of 250K is
dashed, to indicate its likely metastability. In agreement with the
one-dimensional model, the strength of the hydrophobic interaction
increases with increasing temperature, while its range decreases at the
same time: an inverse relation is observed between the strength and
range of the hydrophobic interaction, althoug less prominently in three
dimensions than in one or two. Unlike the one-dimensional model,
the two- and three-dimensional models show oscillations in the mean
force, reflecting the greater geometrical complexity of the lattice in
higher dimensions. In that sense, the curves resemble more closely what
is seen in models in which the solvent is modeled more realistically
\cite{pratt77,paulaitis96}.  In addition to the strong attraction at short
distances, there is now also a local minimum in the potential of depth
about $kT$ at a distance of three lattice spacings in two dimensions
and one of depth about $2kT$ at a distance of two lattice spacings in
three dimensions.

\begin{table}
\caption{The function $p_{11}$ as a function of temperature $T$, for the
two-dimensional (second column) and three-dimensional model (third
column).  The numbers between brackets are the statistical uncertainties
in the last reported digit, obtained from the standard deviation
divided by the square root of the number of independent simulations.}
\begin{tabular}{l|l|l}
T & $p_{11}$ (2D) & $p_{11}$ (3D) \\
\hline
250 & --                       & $6.3324(8)\cdot 10^{-5}$\\
275 & $4.3715(5)\cdot 10^{-5}$ & $3.9119(8)\cdot 10^{-5}$\\
300 & $2.8156(3)\cdot 10^{-5}$ & $2.7108(4)\cdot 10^{-5}$\\
325 & $2.0510(3)\cdot 10^{-5}$ & $2.0177(3)\cdot 10^{-5}$\\
350 & $1.5913(3)\cdot 10^{-5}$ & $1.5791(4)\cdot 10^{-5}$\\
375 & $1.2876(2)\cdot 10^{-5}$ & $1.2822(2)\cdot 10^{-5}$
\end{tabular}
\label{tab:p11}
\end{table}

\begin{figure}
\epsfxsize 8cm
\epsfbox{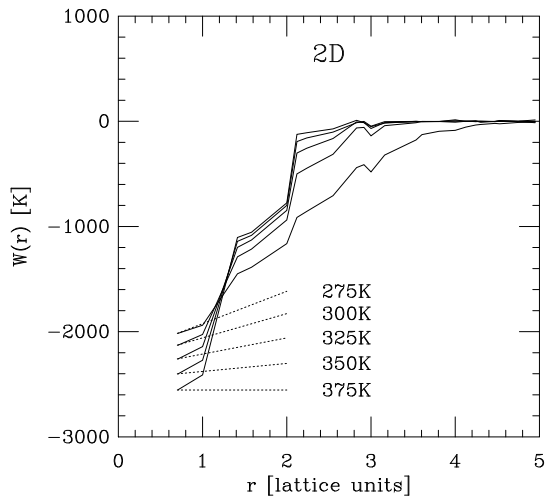}
\caption{The solvent-mediated part of the potential of mean force
$W(r)$, in degrees Kelvin, as a function of distance $r$ between the
solute molecules, in lattice units, for the two-dimensional model. The
statistical error in $W(r)$ is at most two Kelvin, too small to be
plotted.}
\label{twodim}
\end{figure}

\begin{figure}
\epsfxsize 8cm
\epsfbox{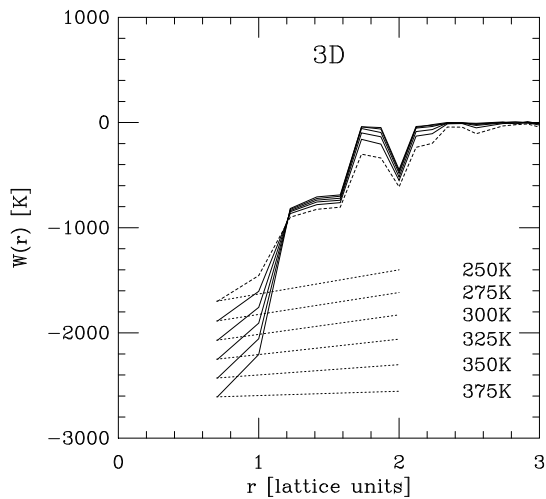}
\caption{The solvent-mediated part of the potential of mean force
$W(r)$, in degrees Kelvin, as a function of distance $r$ between the
solute molecules, in lattice units, for the three-dimensional model.
The statistical error in $W(r)$ is at most two Kelvin, too small to be
plotted. The curve obtained at a temperature of 250K is dashed, to indicate
its likely metastability.}
\label{threedim}
\end{figure}

\section{Summary and conclusions}
\label{sec:summary}

A one-dimensional lattice model of hydrophobic attraction that was
studied recently~\cite{kolomeisky99} is extended to two and three
dimensions. Monte Carlo simulations have been used to determine the
solvent-mediated contribution to the potential of mean force between
hydrophobic solute molecules, and the solubility of the solute.

As in the one-dimensional model, an inverse relation is observed
between the strength and range of the hydrophobic interaction. With
increasing temperature, the strength of the hydrophobic interaction
increases, while at the same time its range decreases, although the
effect is less prominent in three dimensions than in two or one.
Unlike in the one-dimensional model, the force no longer varies
monotonically with distance, but shows oscillations, reflecting the
greater geometrical complexity of the lattice in the higher
dimensions.

At low temperatures, the solubility of the solute is found to decrease
with increasing temperature. This is another signature of the hydrophobic
effect, and also agrees with what had been found in the one-dimensional
model.

\section*{Acknowledgements}

BW is grateful to the Institute of Theoretical Physics and to the Debye
Institute of the University of Utrecht for their hospitality in the
fall of 1999. His work at Cornell was supported by the U.S.  National
Science Foundation and the Cornell Center for Materials Research. The
authors thank M.D. Zeidler for calling their attention to the work in
Reference 2.

\end{document}